\documentclass[rnote]{aa}
\usepackage{graphicx}
\usepackage{amssymb,amsmath}
\usepackage{natbib}
\def\mnras{MNRAS}
\begin{document}
\title{Approximate analytic expressions for
circular orbits around rapidly rotating compact stars} 
\author{M. Bejger \and J. L. Zdunik \and P. Haensel}
\institute{N. Copernicus Astronomical Center, Bartycka
18, PL-00-716 Warszawa, Poland\\
{\tt [bejger;jlz;haensel]@camk.edu.pl}}
\offprints{M. Bejger}
\date{Received 2 August 2010 / Accepted 12 August 2010}
\abstract{}
{We calculate stationary configurations of rapidly rotating compact 
stars in general relativity, to study the properties of circular 
orbits of test particles in the equatorial plane. We search for 
simple, but precise, analytical formulae for the orbital frequency, 
specific angular momentum, and binding energy of a test particle 
that are valid for any equation of state and any rotation frequency 
of the rigidly rotating compact star, up to the mass-shedding limit.}
{Numerical calculations are performed using precise 2-D codes based 
on multi-domain spectral methods. Models of rigidly rotating 
neutron stars and the space-time outside them are calculated for 
several equations of state of dense matter. Calculations are also 
performed for quark stars consisting of self-bound quark matter.}
{At the mass shedding limit, the rotational frequency converges to a 
Schwarzschildian orbital frequency at the equator. We show that the 
orbital frequency of any orbit outside equator can also be 
approximated by a Schwarzschildian formula. Using a simple 
approximation for the frame-dragging term, we obtain approximate 
expressions for the specific angular momentum and specific energy 
on the corotating circular orbits in the equatorial plane of 
neutron star, which are valid to the stellar equator. The formulae 
recover reference numerical values with typically 1\% of accuracy 
for neutron stars with $M\gtrsim 0.5~{\rm M}_\odot$. These values 
are less precise for quark stars consisting of self-bound quark 
matter.}{} 
\keywords{dense matter -- equation of state -- stars: neutron -- stars: rotation}
\titlerunning{Properties of orbits around compact stars}
\authorrunning{Bejger et al.}
\maketitle

\section{Introduction}
\label{sect:intro}
Circular orbits of test particles moving freely along geodesics in a neutron star equatorial plane represent, to a very good
approximation, orbits of gas elements in a low-mass X-ray binary
(LMXB) thin accretion disk. The simplest model of these orbits is
obtained for a Schwarzschild space-time produced by a static
neutron star \citep{ST83}. However, accretion in LMXBs is
associated with neutron star spin-up, and is a commonly accepted
mechanism for producing millisecond pulsars \citep
{AlparCheng1982,BhattaHeuvel1991}. This scenario is corroborated by
 the discovery of rapid pulsations with frequencies up to $619$~Hz, in
more than a dozen LMXBs. List of such bursters is given in Table
1 of \citet{Kiziltan2009}, which already needs an
update: according to \citet{GallowayLCH10}, the transient
burster EXO 0748$-676$ pulsates actually at 552 Hz
instead of 45 Hz, and a new
source Swift J1749.4$-2807$, pulsating at 518 Hz, was reported by
\citet{Altamirano2010}. Nine neutron stars in LMXBs spin at
frequencies ranging from 524 Hz to 619 Hz, and their rotation significantly affects the space-time and thus the orbits of test
particles in their vicinity.

At a given distance from the star center,  a test particle moves at an
orbital frequency $f_{\rm orb}$ on a circular orbit in an
equatorial plane, with the energy and angular
momentum per unit rest mass $\widetilde{E}$ and $\widetilde{l}$,
respectively. Axial symmetry is assumed, and all quantities are
defined in a chosen reference system, using suitable
space-time coordinates.  The star is assumed to be rigidly rotating,
its gravitational mass and angular momentum being $M$ and $J$,
respectively. Analytical formulae expressing $f_{\rm orb}$,
$\widetilde{E}$, and $\widetilde{l}$ as functions of radial
coordinates were obtained in the slow-rotation approximation,
keeping only linear terms in the star's angular momentum $J$, by
\citet{KluzWag1985}. Their expressions coincide, as they should,
after an appropriate change of coordinates, with those obtained in
the lowest order in $J$ for the Kerr metric of a rotating black
hole of the same $M$ and $J$ (see, e.g., \citealt{ST83}). However,
exact 2-D calculations show that the slow-rotation approximation is
not valid   for orbits in the vicinity of compact stars
(neutron stars and quark stars) rotating at
frequency higher than $400$~Hz
\citep{MillerLambCook1998,ShibataS1998,ZdunikHG2002,BertiWMB2005}.

The present research was motivated by a specific astrophysical
project: modeling of the  spin-up process of an old "radio-dead
pulsar", via accretion within a LMXB, into a millisecond pulsar 
\citep{BhattaHeuvel1991}.
During this recycling process, the neutron-star
magnetic field pushes the inner radius of thin Keplerian disk to
$r_0$, which is usually significantly larger than the radius of the
innermost stable circular orbit around an idealized zero magnetic
field star. To perform efficiently a large number of simulations 
corresponding to different astrophysical scenarios, one needs
reliable analytic expressions for $f_{\rm orb}(r_0)$ and
$\widetilde{l}(r_0)$, valid for any rotation frequency up to
the mass-shedding limit $f_{\rm K}$, and for $r_0$ ranging from the
stellar radius to a few thousand of radii. This range of $r_0$ is
required to model the complete process of the recycling of an old
"dead" pulsar rotating at $\sim 0.1$~Hz, with polar magnetic field
$\sim 10^{12}$~G, to a millisecond pulsar rotating at $\sim
500-1000$~Hz, and also a pulsar with a polar magnetic field $\sim 10^{8}-10^9$~G \citep{LorimerLR2008}. 
The inspiration for the present note
came from our previous puzzling result on the implications of 
(still unconfirmed) detection
of neutron star rotation at 1122 Hz for the  equation of state
(EOS) of dense matter \citep{Bejger1122Hz}. At a given
gravitational mass $M$, we found remarkably that the 
orbital frequency in the {\it Schwarzschild}
space-time produced by $M$, calculated at the actual
mass-shedding equatorial circumferential radius, coincided (within
a fraction of a percent) with a {\it true} mass-shedding frequency
for this $M$, which is equal to the {\it true} orbital frequency of
a test particle at the equator \citep{Bejger1122Hz,HaenselZB2008}.
The Schwarzschildian  formula  yielded true  $f_{\rm orb}$,
when one inserted  into it an actual equatorial radius. As we show explicitly in the present note, this is due to the mutual
cancelation of the  effects of both the dragging of the inertial frames 
and the neutron star oblateness.

In the present note, we show that using a Schwarzschildian formula for $f_{\rm orb}(r_0)$, combined with  simple approximations for basic quantities (metric functions, orbital velocity) one can reproduce very precisely (within a fraction of a  percent) orbital parameters {\it outside} a rotating neutron star, down to the star surface, and for rotation frequencies up to the mass-shedding limit. We explain this remarkable property using existing expansions of  $f_{\rm orb}$, $\widetilde{E}$, and $\widetilde{l}$ in the powers of the specific stellar angular momentum \citep{AAKT2003}. In addition, using exact 2-D calculations of space-time of a rotating neutron star, we study how Schwarzschildian $f_{\rm orb}$ at the  equator converges, in the mass-shedding  limit, to the star rigid-rotation frequency, a feature pointed out already in \citet{Bejger1122Hz}.

Notations, space-time coordinates, and exact expressions for the metric functions and  for quantities associated with circular equatorial orbits are introduced in Sect.~\ref{sect:numerical}. Two well-known analytic models of equatorial orbits are summarized in 
Sect.~\ref{sect:approx.analytic}. Our approximation for $f_{\rm orb}$ is described in Sect.~\ref{sect:freq_approx}. Approximations for $\widetilde{l}$  and $\widetilde{E}$ are presented in Sect.~\ref{sect:le_approx}. Approximations for $f_{\rm orb}$, $\widetilde{E}$, and $\widetilde{l}$ are compared with the exact values of these quantities, and their high precision is explained using systematic expansions derived by \citet{AAKT2003}. The formula for the specific energy of test particle, given in \citet{AAKT2003}, contains a misprint that we correct. Finally, Sect.~\ref{sect:discussion} contains a discussion of our results and conclusions. Details concerning equations of state, used in our calculations, are collected in the Appendix.

\begin{figure}[ht]
\resizebox{3.6in}{!}
{\includegraphics[angle=0,clip]{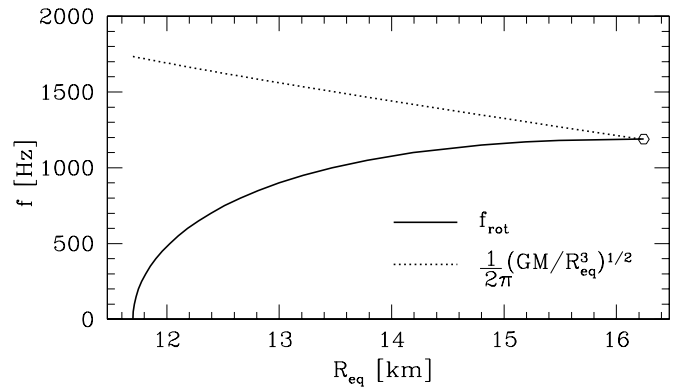}}
\caption{Frequency of stellar rigid rotation (solid line)
compared with the frequency of a test particle orbiting
a point mass, equal to the mass of the rotating star
(Schwarzschildian frequency),  at a distance
of stellar circumferential radius (dotted line). Both frequencies
approximately coincide at the mass-shedding configuration radius 
(marked with a circle). Stellar model based on the DH EOS of \citet{DH2001}
described in the Appendix.}
\label{freqs_surface_comparison}
\end{figure}
\section{Notations, metric, and orbits}
\label{sect:numerical}
The metric for a rapidly rotating object, assuming that
the space-time is axisymmetric, stationary, asymptotically flat, and free of
meridional currents can be written by means of the 3+1 formalism of GR, in the
so-called maximal-slicing quasi-isotropic coordinates
\begin{eqnarray}
ds^2 = -N^2dt^2 &+& A^2(dr^2 + r^2d\theta^2) \nonumber
\\&+& B^2r^2\sin^2\theta(d\phi - N^\phi dt)^2,
\label{metric_numerical}
\end{eqnarray}
where $A$ and $B$ are factors of the 3-metric
($A\equiv B$ in spherical symmetry), $N$ is called the lapse, and $N^\phi$
is the only non-zero component of the shift vector. The integrals of motion
can be obtained by constructing the Lagrangian and using the variational
principle (see e.g., \citealt{ST83}). There are two integrals of motion
of a test particle of rest mass $m$, corresponding to the $t$ and $\phi$
coordinates: $E$, the energy of a test particle (in units of $mc^2$), and
$l$, the specific angular momentum (in units of $mc$). One can also
define $V$, the effective potential
\begin{equation}
V^2 = N^2\left(1 + \frac{l^2}{B^2 r^2}\right)
+ 2N^\phi El - {N^\phi}^2 l^2.
\end{equation}
Conditions for circular orbits are
\begin{equation}
E^2 = V^2~~~ {\rm and}
\label{cond_circular1}
~~~ V_{,r} = 0.
\end{equation}
The radius $r_{circ}$ of the circular orbit in the equatorial plane
of the star ({\it circumferential radius}) is related to the coordinate
radius by the relation
\begin{equation}
r_{circ} = Br .
\label{r_circ}
\end{equation}
The angular velocity of the matter, as measured by
a distant observer, can be described by the components of the
4-velocity vectors $u$ ($g_{\alpha\beta}u^\alpha u^\beta = -1$) and equals
$\Omega=u^\phi/u^t = d\phi/dt$.
Following \citet{Bardeen1970}, the proper velocity of matter
in the equatorial plane is
\begin{equation}
v = (\Omega - N^\phi)\frac{r_{circ}}{N}\cdot
\label{vel_matter}
\end{equation}
The specific (per unit mass) energy $\widetilde{E}$ of a particle
and its specific angular momentum
$\widetilde{l}$ can be expressed in terms of the velocity $v$
\citep{Bardeen1972}:
\begin{equation}
\widetilde{E} = \frac{N + N^\phi vr_{circ}}{\sqrt{1-v^2}},
~~~~~\widetilde{l} = \frac{vr_{circ}}{\sqrt{1-v^2}}.
\label{el_spec}
\end{equation}
From both expresions in Eq.~(\ref{cond_circular1}), one obtains the
{\it corotating test particle} velocity $v$
\begin{eqnarray}
v&=&\Big(\sqrt{4r(\ln N)_{,r}
  + 4r^2(\ln N)_{,r}(\ln B)_{,r}
  +\Big(r^2\frac{B}{N}N^\phi_{,r}\Big)^2}\nonumber \\
  &+& r^2\frac{B}{N}N^\phi_{,r}\Big)/
  \Big(2 + 2r(\ln B)_{,r}\Big),
\end{eqnarray}
and, from Eq.~(\ref{vel_matter}), the orbital frequency
in the equatorial plane
\begin{equation}
f_{orb} = \frac{1}{2\pi}\left( \frac{Nv}{r_{circ}} + N^{\phi}\right) .
\label{num_freq}
\end{equation}
\section{Approximate analytic solutions}
\label{sect:approx.analytic}
\subsection{Schwarzschild solution}
\label{subsect:schwarzschild}
Simplest approximation consists of neglecting neutron star
rotation, and dealing  with the exterior space-time of stationary,
spherically-symmetric object. The metric  in the
Schwarzschild coordinates is then
\begin{eqnarray}
ds^2 = -\left(1 - \frac{r_s}{r}\right)c^2 dt^2
&+& \frac{dr^2}{1 - r_s/r}\nonumber\\
&+& r^2(d\theta^2 + \sin^2\theta d\phi^2),
\label{metric_schwarzschild}
\end{eqnarray}
where $r_s = 2GM/c^2$ is the Schwarzschild radius and $M$ is the
total gravitational mass of the object. The coordinate radius $r$ is
here equal to the cicumferencial radius $r_{circ}$ defined
in Eq.~(\ref{r_circ}). Specific energy $\widetilde{E}^{Schw.}$ and specific
angular momentum $\widetilde{l}^{Schw.}$ are given by the formulae
\begin{equation}
\widetilde{E}^{Schw.} = \frac{r - r_s}
{\left(r^2 - 1.5r_s r\right)^{1/2}},
~~~\widetilde{l}^{Schw.} = \sqrt{\frac{GMr^2}{r - 1.5r_s}}.
\end{equation}
The corresponding orbital frequency $f^{Schw.}_{orb}$ is
\begin{equation}
f^{Schw.}_{orb} = \frac{1}{2\pi}\sqrt{\frac{GM}{r^3}}.
\label{schw_freq}
\end{equation}
\subsection{Slow-rotation approximation}
\label{subsect:sr}
Another popular approximation can be  easily obtained from the Kerr
solution \citep {Kerr1963} in Boyer-Lindquist coordinates; this
{\it slow-rotation} approximation retains only the first-order
terms in the angular momentum $J$ (or equivalently, in $a=J/Mc$).
The formulae for $\widetilde{E}^{sr}$, $\widetilde{l}^{sr}$ and
$f^{sr}_{orb}$ are
\begin{eqnarray}
\widetilde{E}^{sr} &=& \widetilde{E}^{Schw.}\Bigg(1
- a\frac{r^{3/2}_s}{\sqrt{8r}(r - 1.5r_s)(r- r_s)}\Bigg), \\
\widetilde{l}^{sr} &=& \widetilde{l}^{Schw.}
\left(1 - 3a\sqrt{\frac{r_s}{2r^3}}\frac{r-r_s}{r-1.5r_s}\right),\\
f^{sr}_{orb} &=& f^{Schw.}_{orb}
\left( 1 - a\sqrt{\frac{r_s}{2r^3}} \right).
\label{sr_freq}
\end{eqnarray}
\section{Approximation of orbital frequency by the Schwarzschildian
 formula}
\label{sect:freq_approx}
The seemingly surprising property of angular frequency $f_{orb}$
being approximately equal to $f^{Schw.}_{orb}$ can be explained by
studying the higher-order expansion terms of axisymmetric metrics, such as
Hartle-Thorne metric, presented by \citet{AAKT2003} (hereafter referred to as AAKT). AAKT give the expansion to the second order in angular momentum and include the mass-quadrupole moment, $Q$, term, which
plays a decisive r{\^o}le in a correct
description of the orbital parameters of a test particle
\citep{ShibataS1998,LaarakkersP1999,BertiWMB2005}.
The {\it orbital angular velocity} of a corotating
test particle can be approximated, according to AAKT, as
\begin{equation}
\Omega = \frac{M^{1/2}}{r^{3/2}}\Bigg[1 - j\frac{M^{3/2}}{r^{3/2}}
+ j^2F_1(r) + qF_2(r)\Bigg],
\label{aakt_omega}
\end{equation}
where $G=c=1$, $j=a/M$, $q=Q/M^3$ and the functions $F_1(r)$ and $F_2(r)$
are defined therein.

\begin{figure}[ht]
\resizebox{3.6in}{!}
{\includegraphics[clip]{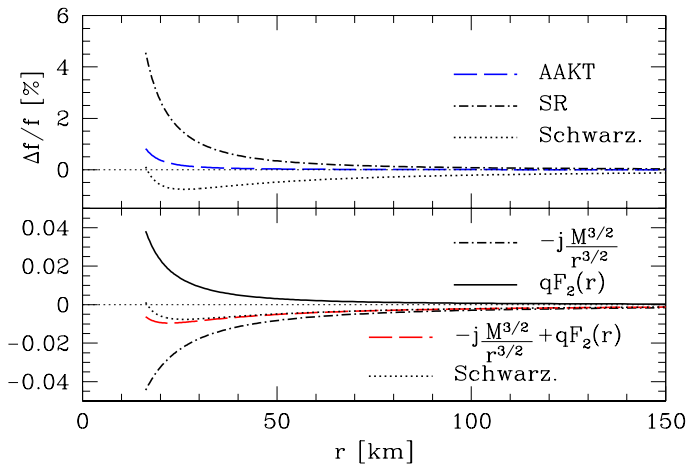}}
\caption{Upper panel: percentage differences between exact numerical
values of test particle orbital frequencies, $f_{\rm orb}$,
and approximations presented in the text, plotted against the
radius of a circular orbit; $\Delta f/f\equiv(f_{orb} - f)/f$, where
 $f = f^{AAKT}_{orb}, f^{sr}_{orb}$, 
or $f^{Schw.}_{orb}$. Lower panel: values of various terms
in Eq.~(\ref{aakt_omega}) versus orbital radius. For comparison,
relative frequency difference in the case of the Schwarzschildian formula
is also shown (mass-shedding configuration of DH EOS, details in Appendix).}
\label{freq_comparison}
\end{figure}

As an illustration, 
we present\footnote{As in \citet{BertiWMB2005} (Sect. 3.1 therein), 
we use {\it exact numerical} values of the gravitational mass $M$, total
angular momentum $J$, equatorial radius $R$, {\it but also} the mass-quadrupole moment $Q$ of a given configuration as input to AAKT expansions to reproduce as well as possible the test particle orbital frequencies, angular momenta, and energies and
compare them with their analogues from different space-times.}
 in Fig.\;2 a comparison of the frequency $f^{AAKT}_{orb}=\Omega/2\pi$,
for the mass-shedding configuration calculated for
realistic EOS of \citet{DH2001}(details in the Appendix),  with $f_{orb}$,
$f^{Schw.}_{orb}$, and $f^{sr}_{orb}$ (Eqs.~\ref{num_freq},~\ref
{schw_freq}, and \ref{sr_freq}, respectively). As expected,
$f^{AAKT}_{orb}$ reproduces the true, numerically-obtained values
of $f_{orb}$ far more precisely than $f^{sr}_{orb}$, especially
near the stellar surface. One has to keep in mind however that in
Eq.~(\ref {aakt_omega}) we have used the mass-quadrupole moment
obtained directly from our numerical calculations instead of an
approximation to $q$, which may affect the behavior of
$f^{AAKT}_{orb}$. The value of $f^{AAKT}_{orb}$ is also quite close to
$f^{Schw.}_{orb}$. We sought for an explanation of this
phenomenon by analyzing the terms in Eq.~(\ref{aakt_omega}): in
addition to them being small in comparison to the leading term, the
$j$-term is {\it approximately} equal, but with the opposite sign,
to the mass-quadrupole moment $q$-term, so that they effectively 
cancel each other (see Fig.~\ref{freq_comparison}
for an typical comparison; the $j^2$-term is much smaller than both
of the other terms, two orders of magnitude in this example, and
therefore insignificant). This feature is {\it qualitatively} and
{\it quantitatively} present for different polytropic and realistic
EOSs as well as bare strange-quark matter stars (details concerning
the EOSs that we used can be found in the Appendix).

The quadrupole term in Eq.\;(\ref{aakt_omega}) is related to
the rotational oblateness of the star and, in contrast to $j$-terms,
is of Newtonian nature. One can study the maximum deviation from
the Schwarzschildian (and Newtonian) test particle orbital frequency
resulting from this effect, assuming a dense matter
disk (which corresponds to an "extreme oblateness")
as a source of gravitational field. The gravitational pull is described
by hypergeometric functions \citep{ZduGou2001} that indicate the maximum
frequency deviation $\Delta f/f\simeq 18$\% for orbital radii
larger than that of the innermost stable circular orbit - 
the existence of unstable circular orbits is a result of the 
oblateness of the gravitational field source, treated in the Newtonian theory.
\section{Approximation of the specific orbital energy and angular momentum}
\label{sect:le_approx}
We approximate the specific energy and
angular momentum of a test particle, in a manner similar to
\citet{ShibataS1998}. As one can see in an example shown
in Fig.~\ref{freq_comparison},
the difference between the exact orbital frequency and the
Schwarzschildian  formula is smaller than 1\% even for mass-shedding, compact
stellar configurations. Near  the surface of the star, 
$f^{Schw.}_{orb}$ is far more accurate than the slow-rotation 
approximation, and as good as $f^{AAKT}_{orb}$.

\begin{figure}[ht]
\resizebox{3.6in}{!}
{\includegraphics[clip]{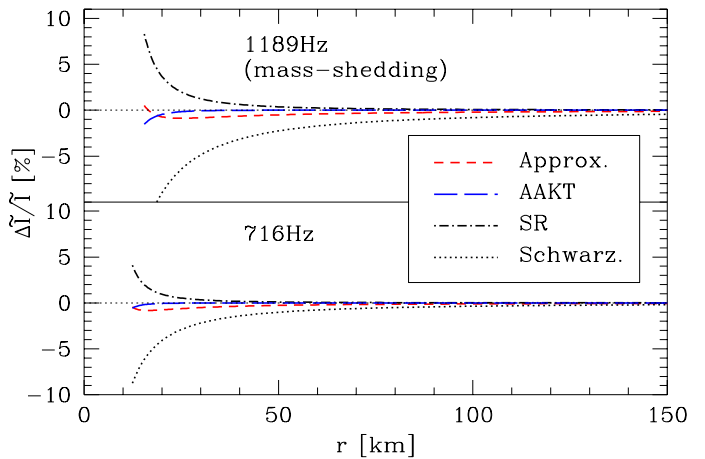}}
\caption{Comparison of specific orbital angular momenta
as a function of the radius of the circular orbit.
The exact value is denoted by $\widetilde{l}$, while
$\widetilde{l}_{\rm a}$ is an approximation of it. The
relative deviations (in percent) are $\Delta\widetilde{l}/\widetilde{l}\equiv(\widetilde{l} -
\widetilde{l}_{\rm a})/\widetilde{l}$,
where $\widetilde{l}_{\rm a}
 = \widetilde{l}^{appr.}, \widetilde{l}^{AAKT}, \widetilde{l}^{sr}$
or $\widetilde{l}^{Schw.}$.
Results for DH EOS mass-shedding configuration and for the
currently-highest pulsar frequency, 716 Hz (details in Appendix).}
\label{fig_dh_lspec_ms_716Hz}
\end{figure}

\begin{figure}[ht]
\resizebox{3.6in}{!}
{\includegraphics[clip]{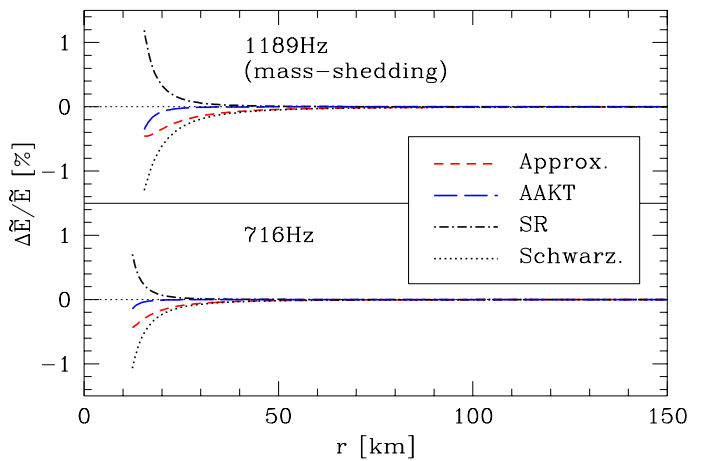}}
\caption{Comparison of specific orbital energies
as a function of the radius of the circular orbit.
Configurations and notation analogous to
Fig.~\ref{fig_dh_lspec_ms_716Hz}.}
\label{fig_dh_espec_ms_716Hz}
\end{figure}

Motivated by this result, we aim to provide formulae that are simpler
than those of AAKT and which could serve the ambition of being
useful in practical calculations. To this aim, we propose to substitute the
exact values for the test particle velocity in the Eq.~(\ref{vel_matter})
by their approximations of the angular frequency $\Omega$ by
$2\pi f^{Schw.}_{orb}$ from Eq.~(\ref{schw_freq}), the azimuthal shift
component $N^\phi$ by the first-order term in the slow-rotation
approximation of a frame-dragging term $2GJ/r^3c^2$, and the metric function
$N$ by its Schwarzschildian $c\sqrt{1 - r_s/r}$:
\begin{equation}
v_{appr.} = (2\pi f^{Schw.}_{orb} - 2GJ/r^3c^2)\frac{r}{c\sqrt{1 - r_s/r}},
\label{vel_matter_approx}
\end{equation}
where $r\equiv r_{circ}$ for brevity. We then apply $v_{appr.}$, as well
as the approximations described above, to Eqs.~(\ref{el_spec}), obtaining
$\widetilde{l}^{appr.}$ and $\widetilde{E}^{appr.}$.
In Figs.~\ref{fig_dh_lspec_ms_716Hz}, and \ref{fig_dh_espec_ms_716Hz} we
compare\footnote{Note the difference in definitions
of the specific orbital angular momentum between the article of \citet{AAKT2003}
and here: $\widetilde{l}^{AAKT} = - u_\phi/u_t = -\widetilde{l}/\widetilde{E}$. We
also notice that there is a misprint (?)
in the Eq.~(30) of AAKT. Their expression leads to divergence, with increasing $j$, of approximate $\widetilde{E}(r)$ from the exact values of this quantity. This could be repaired  by replacing $-20M^2r$ in their Eq.\;(30) by $+20M^2r$.}
our  approximation with the Schwarzschild, slow-rotation, and
AAKT (Eqs.\;(21) and (26) therein) formulae for $\widetilde{l}$ and
$\widetilde{E}$.
For different EOSs, masses, and rates of rotation, the behaviour
of these deviations is quantitatively and qualitatively similar.

\section{Discussion and conclusions}
\label{sect:discussion}
We have performed calculations of stationary configurations of rotating
compact stars, for a set of representative EOSs, both
polytropic and  realistic ones, as well as for the bag models describing the
EOSs of hypothetical bare strange quark stars (as in \citealt{HaenselZBL09}). In all these cases, we observed quantitatively and qualitatively similar behaviour of
the orbital frequency of a test particle moving on a circular orbit in the
equatorial plane. In all cases, the value of the orbital frequency
 was close to that of a test particle in a Schwarzschildian space-time
 around a point mass (equal to that of the actual compact star),  at the
 properly  defined radius. This result is valid for any EOS and for
any rotation frequency of the compact star, up to the mass-shedding limit.

\begin{figure}[ht]
\resizebox{3.6in}{!}
{\includegraphics[clip]{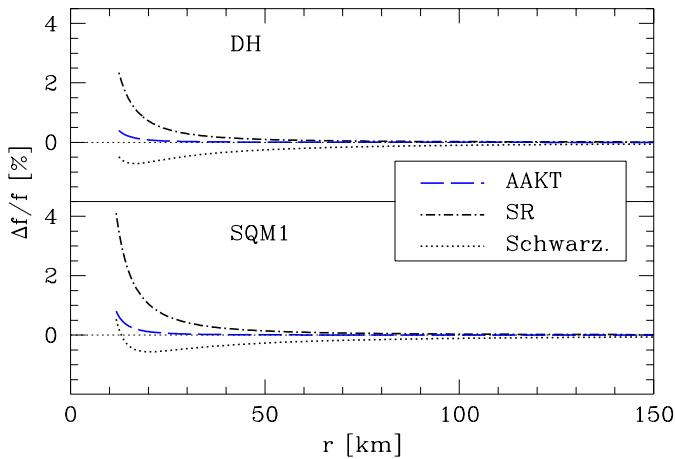}}
\caption{Comparison of orbital frequencies for stellar configurations
of DH EOS and bare strange quark star EOS, rotating at 716 Hz.
Notations as in Fig.~\ref{freq_comparison}, details in Appendix.}
\label{fig_dh_sqm_freq_716Hz}
\end{figure}

We note that the approximations of $\widetilde{l}$ and 
$\widetilde{E}$, proposed in Sect.~\ref{sect:le_approx}, reproduce 
true (numerical) values within  about one per cent (being the least 
accurate in the vicinity of the stellar surface). Their accuracy 
near the stellar surface decreases to a few percent for rapidly 
rotating stars with extreme compactnesses ($2GM/Rc^2\gtrsim 0.5$), 
especially in the case of rapidly rotating and oblate bare 
strange-quark matter stars with large quadrupole moments, but as is 
shown in Fig.~\ref{fig_dh_sqm_freq_716Hz} the approximations are 
acceptable for the currently highest observed spin rate of $716$
~Hz. For comparison, Figs. \ref{fig_dh_lspec_ms_716Hz} and \ref 
{fig_dh_espec_ms_716Hz} show the difference in the  precision for 
the DH EOS in the case of the mass-shedding limit and the rotation 
frequency $716$~Hz.

\begin{figure}[ht]
\resizebox{3.6in}{!}
{\includegraphics[clip]{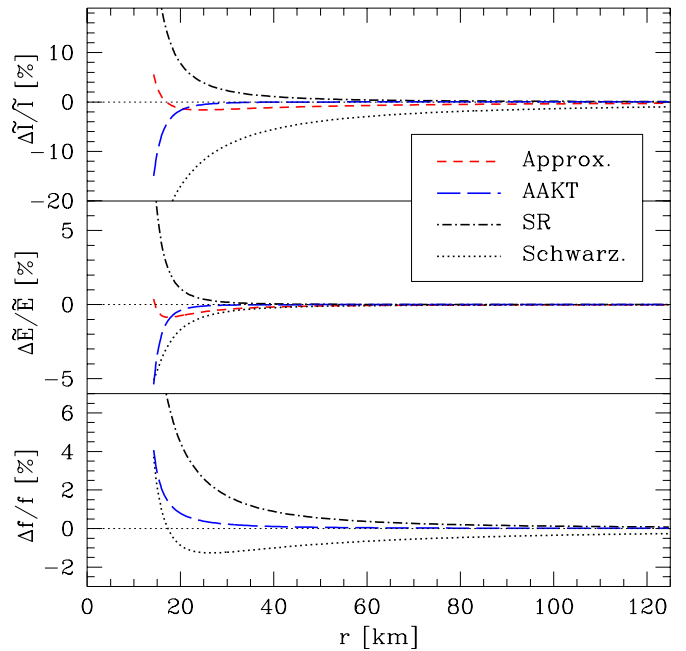}}
\caption{Comparison of $\widetilde{l}$, $\widetilde{E}$, and $f_{orb}$ for
$2.2~M_\odot$ SQM1 EOS star rotating at 1300 Hz. Notation as in previous
figures.}
\label{fig_sqm_lspec_espec_freq_1300Hz}
\end{figure}

Overall, our approximations for $\widetilde{l}$ and $\widetilde{E}$ 
are much better than the slow-rotation approximation, and one does 
not need  to compute the mass-quadrupole moment of the star, the 
knowledge of which is otherwise crucial in the existing systematic 
expansions \citep{BertiWMB2005,AAKT2003}. In cases of larger 
compactness and oblateness, as well as sub-millisecond rotation 
periods, the approximation from Sect.~\ref{sect:le_approx} 
is in closer agreement with the results of numerical simulations than the 
formulae of \citet{AAKT2003}; in Fig.~\ref
{fig_sqm_lspec_espec_freq_1300Hz}, we present an extreme example of a bare 
quark star spinning at 1300Hz.

The analytic formulae for  $f_{\rm orb}(r)$, $\widetilde{l}(r)$, and
$\widetilde{E}(r)$ give a very good approximation (typically within
a one per cent) of exact values for neutron stars with astrophysically
relevant masses, $M\gtrsim 0.5\;{\rm M}_\odot$. They
are valid for rigidly rotating neutron stars, for rotation rates
up to the mass shedding limit, and down to the equator. However, they cannot
be used to calculate the properties of the innermost stable circular orbit,
because this calculation involves second radial derivatives
of metric functions (see, e.g., \citealt{CST1994}).
\acknowledgements
We would like to thank Prof. M. Abramowicz and Prof. P. Jaranowski for 
useful discussions. This work was partially supported by the Polish MNiSW
research grant no. N N203 512838. MB acknowledges the support
of Marie Curie Fellowship no. ERG-2007-224793 within the 7th European
Community Framework Programme.
\section*{Appendix: Numerical implementation and the EOSs chosen for calculations}
\label{sect:appendix}
The calculations have been performed using the {\tt rotstar} code from the
LORENE library\footnote{See {\tt http://www.lorene.obspm.fr} (detailed description of the EOS implementation can be found at {\tt http://www.lorene.obspm.fr/Refquide/classEos.html}).}. 
We use EOSs employed in \citet{HaenselZBL09}:
\begin{enumerate}
\item {\it Realistic microphysical EOSs of dense matter}.
Unless marked otherwise, the figures employ the results for DH EOS of \citet{DH2001},
for a stellar configuration corresponding to a
non-rotating star of mass $M=1.43~M_\odot$ (central baryon density
$n_{b,c} = 0.55~{\rm fm^{-3}}$). The mass-shedding configuration
 has a gravitational mass of $1.8~M_\odot$, whereas that rotating
 at 716 Hz has a mass of $1.52~M_\odot$. 
\item {\it Bag models of bare strange quark stars}.
Figs.~\ref{fig_dh_sqm_freq_716Hz} and \ref{fig_sqm_lspec_espec_freq_1300Hz} show results for SQM1 EOS. 
The non-rotating counterpart to the stellar configuration presented 
in Fig.~\ref{fig_dh_sqm_freq_716Hz} has the gravitational mass of $1.43~M_\odot$. The Fig.~\ref{fig_sqm_lspec_espec_freq_1300Hz}
configuration has a central baryon density of $n_{b,c} = 0.63~{\rm fm^{-3}}$,
non-rotating gravitational mass of $M = 1.7~M_\odot$, and compactness $2GM/Rc^2 = 0.47$.
\item {\it Relativistic polytropes}. We considered the range 
$\gamma = 1.5 - 3$ \citep{Tooper1965}.
\end{enumerate}


\begin{thebibliography}{} {
\bibitem[Abramowicz et al.(2003)]{AAKT2003}
Abramowicz, M. A., et al., 
2003, {\tt arXiv:gr-qc/0312070} (AAKT)

\bibitem[Alpar et al.(1982)]{AlparCheng1982}
Alpar, M.A., et al., 
1982, Nature, 728

\bibitem[Altamirano et al.(2010)]{Altamirano2010}
Altamirano, D., et al., 2010, Astr. Tel., 2565

\bibitem[Bardeen(1970)]{Bardeen1970}
Bardeen, J. M., 1970, ApJ 162, 71

\bibitem[Bardeen(1972)]{Bardeen1972}
Bardeen, J. M., 1972, ApJ 178, 347

\bibitem[Bejger et al.(2007)]{Bejger1122Hz}
Bejger, M., Haensel, P., Zdunik, J.L., 2007,
A\&A  464, L49

\bibitem[Berti et al.(2005)]{BertiWMB2005}
Berti, E., et al., 
2005, \mnras, 358, 923

\bibitem[Bhattacharya \& van den Heuvel(1991)]{BhattaHeuvel1991}
Bhattacharya, D., van den Heuvel, E.P.J., 1991,
Phys. Rep. 203, 1

\bibitem[Cook et al.(1994)]{CST1994}
Cook, G.B., Shapiro, S.L., Teukolsky, S.A., 1994,
ApJ 424, 823

\bibitem[Douchin \& Haensel(2001)]{DH2001}
Douchin, F., Haensel, P., 2001, A\& A, 380, 151

\bibitem[Haensel et al.(2008)]{HaenselZB2008}
Haensel, P., Zdunik, J.L., Bejger, M., 2008,
New Astron. Rev. 51, 785

\bibitem[Haensel et al.(2009)]{HaenselZBL09}
Haensel, P., et al., 
2009 A\& A, 502, 605

\bibitem[Galloway et al.(2009)]{GallowayLCH10}
Galloway, D. K., Lin, J., Chakrabarty, D., Hartman, J. M.,
2010 ApJ, 711, L148

\bibitem[Kerr(1963)]{Kerr1963}
Kerr, R .P., 1963, PRL 11, 237

\bibitem[Kiziltan \& Thorsett(2009)]{Kiziltan2009}
Kiziltan, B., Thorsett, S.E., 2009,
ApJ 693, L109

\bibitem[Klu{\'z}niak \& Wagoner(1985)]{KluzWag1985}
Klu{\'z}niak, W., Wagoner, R.V., 1985,
ApJ 297, 548

\bibitem[Laarakkers \& Poisson(1999)]{LaarakkersP1999}
Laarakkers, W. G., Poisson, E.,
ApJ, 512, 282

\bibitem[Lorimer(2008)]{LorimerLR2008}
Lorimer, D.R., 2008, Living. Rev. Relativity 11, 8. [Online Article]:
cited [20 April 2010], http://www.livingreviews.org/lrr-2008-8

\bibitem[Miller et al.(1998)]{MillerLambCook1998}
Miller, M.C., Lamb, F.K., Cook, G.B., 1998,
ApJ 509, 793

\bibitem[Shapiro \& Teukolsky(1983)]{ST83}
Shapiro, S.~L., \& Teukolsky, S.~A.\ 1983, Wiley-Interscience

\bibitem[Shibata \& Sasaki(1998)]{ShibataS1998}
Shibata, M., \& Sasaki, M.\ 1998, PRD 58, 104011

\bibitem[Tooper(1965)]{Tooper1965}
Tooper, R. F., 1965, ApJ 142, 1541

\bibitem[Zdunik \&  Gourgoulhon(2001)]{ZduGou2001}
Zdunik, J.L., Gourgoulhon, E., 2001,
PRD 63, 087501

\bibitem[Zdunik et al.(2002)]{ZdunikHG2002}
Zdunik, J.L., Haensel, P., Gourgoulhon, E., 2002,
A\&A 381, 933

}
\end{thebibliography}
\end{document}